\newcommand{\rem}[1]{}

\newcommand{\Id}[1] {\int \! \! {\rm d}^3 #1}

\newcommand{\vr} {{\bf r}}

\documentclass[a4paper,12pt]{article}
\usepackage{hyperref}
\begin{document}

\title{
Jahn-Teller distortions and excitation energies in C$_{60}^{n+}$
}

\author{Martin L{\"u}ders$^{1,2}$\thanks{E-mail: lueders@sissa.it} 
        and
        Nicola Manini$^{1,3,4}$\thanks{E-mail: nicola.manini@mi.infm.it}
\\ \small \it 
$^1$ International School for Advanced Studies (SISSA),
\\ \small \it 
Via Beirut 4, 34014 Trieste, Italy
\\ \small \it 
$^2$ INFM Democritos National Simulation Center, 
\\ \small \it 
and INFM, Unit\`a Trieste, Italy
\\ \small \it 
$^3$ Dip.\ Fisica, Universit\`a di Milano, Via Celoria 16, 20133
Milano, Italy
\\ \small \it 
$^4$ INFM, Unit\`a di Milano, Milano, Italy
}

\date{}
\maketitle

\begin{abstract}
Based on previously computed parameters for the electron-phonon couplings
and the Coulomb exchange, we compute and classify the static Jahn-Teller
distortions, i.e.\ the minima of the lowest adiabatic potential energy
surface, of C$_{60}^{n+}$, for all values of charge $1\leq n\leq 9$ and
spin.
We compute the intra-band electronic excitation energies in the different
optimal geometries in the  sudden approximation, and find a spread of the
electronic states of roughly 1~eV.
We also obtain the leading vibronic quantum corrections to the ground-state
energy, equal to zero-point energy lowering due to the softening of the
phonons at the adiabatic Jahn-Teller minima: these non-adiabatic
corrections are so large that for $4\leq n\leq 6$ states of different
spin symmetry turn lower than the high-spin adiabatic ground state.
\end{abstract}

\thispagestyle{empty}
\newpage

\section{Introduction}

Low-spin states are associated to larger distortions, thus larger energy
gains, than high-spin states in degenerate electron-phonon coupled
molecules and impurity centers.
Electron-electron Coulomb repulsion opposes this tendency, favoring
high-spin states instead, in accord to the first of Hund's rules.
The Jahn-Teller (JT) systems C$_{60}^{n+}$ are no exception to this rule:
if electron-phonon coupling was the only relevant interaction, then the
$n$-holes ground state would be either of spin $S=0$ (even $n$) or 
$S=\frac 12$ (odd $n$).
As was recently shown \cite{Lueders02}, in positive fullerene ions
the size of Coulomb interaction is sufficiently large to enforce Hund's
rule: the ground states of C$_{60}^{n+}$ was calculated to always be high spin
($S=\frac n2$ for $n\leq 5$, $S=\frac {10-n}2$ for $n>5$) in the adiabatic
approximation.
This result is confirmed for $n=2$ by NMR investigation of solid-state
compounds \cite{Panich02}.
The JT distortions in C$_{60}^{n+}$, though strongly counteracted by the
larger electron-electron repulsion, yet represent an important, and still
largely unexplored, contribution to the energetics of C$_{60}^{n+}$.
Investigation of this contribution, and in particular of the corrections to
the adiabatic approximation, is the main subject of this work.

The JT model relevant for C$_{60}^{n+}$ is conventionally indicated as $h^n
\otimes (A + G + H)$, where $h$ refers to the fivefold-degenerate highest
occupied molecular orbital (HOMO), and $A$, $G$, $H$ refer to the 2
nondegenerate $A_g$, 6 fourfold-degenerate $G_g$ and 8 fivefold-degenerate
$H_g$ molecular vibration modes that are linearly coupled to the $h_u$
states according to icosahedral symmetry \cite{Lueders02,CeulemansII,hbyh}.
We investigate this model by treating the normal coordinates for these
vibrational modes as classical variables, and searching the minima of the
adiabatic potential energy surface in the 66-fold dimensional space of
these distortions.
Each of these static JT configurations is characterized by a reduced
symmetry from icosahedral to some (usually) lower symmetry.
New vibrational frequencies arise at these local minima: we determine these
frequencies by evaluation of the Hessian matrix at the minimum
\cite{ManyModes}.
The lowering of the vibrational frequencies gives the leading quantum
correction to the adiabatic approximation.
The original icosahedral symmetry of the problem is restored once the
presence of several equivalent optimal distortions is recognized, and
quantum tunneling between these wells is allowed.
Proper accounting of tunneling gives the next-order quantum correction, but
in the present work, we limit ourselves to the study of the local
properties of the wells and the connectivity of the sets of minima in
distortion space, for all values of charge $n$ and spin $S$.

The competing intra-molecular exchange of Coulomb origin and the JT
interaction both contribute to the computed spectrum of excitations.
Differences in energies of the fully relaxed configurations at different
spin compare directly with spin gaps as could be measured in ``slow''
spectroscopies such as electron or nuclear magnetic resonance.
In contrast, the electronic excitation energies computed keeping the
molecular geometry fixed in the lowest minimum compare directly with the
vertical excitations probed by fast optical spectroscopies.
Both these class of quantities are reported in this work.

This paper is organized as follows: Sec.~\ref{model:sec} introduces the
model and the parameters used in this calculation, which is then described
in Sec.~\ref{adiabatic:sec}, along with the properties of the JT minima for
all values $n$ and $S$; Sect.~\ref{excitations:sec} contains the vertical
excitation spectra.
The the zero-point non-adiabatic corrections are described in
Sec.~\ref{nonadiabatic:sec}.
The results are discussed in Sec.~\ref{discussion:sec}, and connectivity
matrices are collected in an Appendix.

\section{The model Hamiltonian}

\label{model:sec}

We report here for completeness the model Hamiltonian previously introduced
in Ref.~\cite{Lueders02} to describe the physics of the holes in the $h_u$
HOMO of C$_{60}$ fullerene:
\begin{equation}
\hat{H} = \hat{H}_0 + \hat{H}_{\rm vib}  + 
\hat{H}_{\rm e-v} + \hat{H}_{\rm  e-e}
\label{modelhamiltonian}
\end{equation}
where 
\begin{eqnarray}
\hat{H}_0     &=& \epsilon \, \sum_{\sigma m}  
\hat{c}^\dagger_{\sigma m} \hat{c}_{\sigma m} \\
\label{vib-hamiltonian}
\hat{H}_{\rm vib} &=& \sum_{i\Lambda \mu} \frac{\hbar \omega_{i\Lambda}}{2} 
( \hat{P}_{i\Lambda \mu}^2 + \hat{Q}_{i\Lambda \mu}^2 ) \\
\label{JT-hamiltonian}
\hat{H}_{\rm e-v} &=& 
\sum_{r\,i\Lambda}
\frac{k^\Lambda g^r_{i\Lambda} \hbar \omega_{i\Lambda}}{2} 
\sum_{\sigma m m' \mu} 
C^{r \Lambda \mu}_{m m'} \, \hat{Q}_{i\Lambda \mu} \,\hat{c}^\dagger_{\sigma m}
\hat{c}_{\sigma m'} \\
\hat{H}_{\rm  e-e} &=& 
\frac{1}{2} \sum_{\sigma, \sigma'} \sum_{{m m'}\atop{n n'}} 
w_{\sigma,\sigma'}(m,m';n,n') \,
\hat{c}^\dagger_{\sigma m} \hat{c}^\dagger_{\sigma' m'} \, 
\hat{c}_{\sigma' n'} \hat{c}_{\sigma n}.
\label{Coulomb-hamiltonian}
\end{eqnarray}
are respectively the single-particle Hamiltonian, the vibron contribution
(representing the phonon kinetic energy plus the restoring potential
expanded to quadratic order around the equilibrium configuration of neutral 
C$_{60}$), the electron-vibron coupling (in the linear JT approximation)
\cite{hbyh,Manini01}, and finally the mutual Coulomb repulsion between the
electrons.
The $\hat{c}^\dagger_{\sigma, m}$ denote the creation operators of a hole
in the HOMO, described by the single-particle wave function
$\varphi_{m\sigma}(\vr)$.
$\sigma$ indicates the spin projection; $m$ and $n$ label the component
within the fivefold degenerate electronic HOMO multiplet, based on the
$C_5$ quantum number $m$ from the $I_h\supset D_5\supset C_5$ group chain
\cite{hbyh,Butler81}.
$i$ counts the phonon modes of symmetry $\Lambda$ (2 $A_g$, 6 $G_g$ and 8
$H_g$ modes).
$C^{r \Lambda \mu}_{m n}$ are Clebsch-Gordan coefficients \cite{Butler81}
of the icosahedral group $I_h$, for coupling two $h_u$ states to phonons of
symmetry $\Lambda$.
$r$ is a multiplicity label, relevant for modes of $\Lambda=H_g$ symmetry
only \cite{Manini01,Butler81}.
$\hat{Q}_{i\Lambda \mu}$ are the molecular normal-mode vibration
coordinates (measured from the adiabatic equilibrium configuration of
C$_{60}$), and $\hat{P}_{i\Lambda \mu}$ the corresponding conjugate
momenta.
Spin-orbit is exceedingly small in C$_{60}$ \cite{TMG} and it is therefore
neglected.

\begin{table}
\begin{center}
\begin{tabular}{rrcrrr}
\hline
\hline
$\hbar\omega_{\tau i}$  &       $\hbar\omega_{\tau i}$  &
$g_{\tau i}$    &       $\alpha_{\tau i}$       
&       $E_{\rm s}(D_5)$        &$E_{\rm s}(D_3)$       \\
cm$^{-1}$               &       meV     &       &  deg
&       meV     &       meV \\
\hline
$A_g$\quad\quad\quad\\
500     &       62.0    &       0.059   &       -       &       0.0     &       0.0     \\
1511    &      187.4    &       0.274   &       -       &       1.8     &       1.8     \\
\hline
$G_g$\quad\quad\quad\\
483     &       59.9    &       0.757   &       -       &       0.0     &       1.9     \\
567     &       70.3    &       0.102   &       -       &       0.0     &       0.0     \\
772     &       95.7    &       0.800   &       -       &       0.0     &       3.4     \\
1111    &      137.8    &       0.624   &       -       &       0.0     &       3.0     \\
1322    &      163.9    &       0.228   &       -       &       0.0     &       0.5     \\
1519    &      188.4    &       0.467   &       -       &       0.0     &       2.3     \\
\hline
$H_g$\quad\quad\quad\\
261     &      32.4     &       3.042   &       $-0.1$  &      30.0     &       0.0     \\
429     &      53.2     &       1.223   &       30.1    &       6.0     &       1.1     \\
718     &      89.0     &       0.995   &       89.4    &       0.0     &       4.9     \\
785     &      97.3     &       0.784   &       $-2.3$  &       6.0     &       0.0     \\
1119    &     138.7     &       0.221   &       76.6    &       0.0     &       0.4     \\
1275    &     158.0     &       0.519   &       28.0    &       3.3     &       0.5     \\
1456    &     180.5     &       0.962   &       28.1    &      13.0     &       2.1     \\
1588    &     196.9     &       0.869   &     $-31.1$   &      10.9    &       2.2     \\
\hline
\hline
\end{tabular}
\end{center}
\caption{Computed mode eigenfrequencies and e-v linear coupling parameters
of the $h_u$ HOMO in C$_{60}$ \cite{Manini01}.  The classical single-mode
JT stabilization energies $E_{\rm s}$ are tabulated for both $D_5$ and
$D_3$ distortions, for {\em one hole} in the HOMO.
\label{e-v-params:table}
}
\end{table}

\begin{table}
\begin{center}
\begin{tabular}{cr}
\hline
\hline
Parameter & Value\\
        &       [meV]  \\
\hline
$F_1$   &15646 $\pm$  9 \\
$F_2$   &  105 $\pm$ 10 \\
$F_3$   &  155 $\pm$  4 \\
$F_4$   &   47 $\pm$  5 \\
$F_5$   &    0 $\pm$  3 \\
\hline
$U$     & 3097 $\pm$  1 \\
\hline
\hline
\end{tabular}
\caption{
The Coulomb parameters for C$_{60}^{n+}$, as obtained from the DF
calculations of Ref.~\cite{Lueders02}.
One of the tabulated parameters (e.g.\ $F_1$) is a linear
combination of the five others.
\label{e-e-params:table}
}
\end{center}
\end{table}

The electron-vibron  (e-v) couplings $g^r_{i\Lambda}$ are conveniently
expressed in units of the corresponding harmonic vibron quantum of energy
$\hbar\omega_{i\Lambda}$.
In this calculation we adopt the numerical values of the e-v coupling
parameters, listed in Table~\ref{e-v-params:table}, from the Density
Functional (DF) calculation of Ref.~\cite{Manini01}, and
a second calculation \cite{Saito02} yields couplings in substantial accord
with those of Table~\ref{e-v-params:table}.
The numerical factors $k^{A_g}=5^{\frac 12}$, $k^{G_g}=\left(\frac 54
\right)^{\frac 12}$, $k^{H_g}=1$ in $\hat{H}_{\rm e-v}$ have been
introduced for compatibility with the normalization of
Ref.~\cite{Manini01}.

The Coulomb matrix elements are defined by:
\begin{equation}
\label{Coulomb-ints}
w_{\sigma,\sigma'}(m,m';n,n') = \Id{r} \! \Id{r'} \, 
\varphi^{*}_{m \sigma}(\vr) \, 
\varphi^{*}_{m'\sigma'}(\vr') \,
u_{\sigma,\sigma'}(\vr,\vr') \, 
\varphi_{n\sigma}(\vr) \,
\varphi_{n'\sigma'}(\vr') 
\end{equation}
where $u_{\sigma,\sigma'}(\vr,\vr')$ is an effective Coulomb repulsion,
screened by the other electrons of the molecule.
Detailed symmetry analysis shows \cite{Lueders02} that, assuming
spin-independence of the orbitals, this set of
coefficients can be expressed as 
\begin{equation}
w_{\sigma,\sigma'}(m,m';n,n') = \sum_{r,r',\Lambda} F^{r,r',\Lambda}
\left( \sum_\mu C^{r \Lambda \mu}_{m n} \, 
C^{r' \Lambda \mu}_{m' n'} \right) 
\label{Fdecomposition}
\end{equation}
in terms of a {\em minimal} set of independent parameters
$F^{r,r',\Lambda}$.
A DF calculation of these parameters was carried out in
Ref.~\cite{Lueders02}, and for our calculation we adopt those values of
the Coulomb parameters, which we report for completeness in
Table~\ref{e-e-params:table}.
For the Coulomb parameters we use the shorthands
\begin{equation}
F_1= F^{A_g},\
F_2= F^{G_g},\
F_3= F^{1,1,H_g},\
F_4= F^{2,2,H_g},\
F_5= F^{1,2,H_g},
\label{fnumbering}
\end{equation}
and the combination
\begin{equation}
U = \left( \frac{F_1}{5} - \frac{4 \, F_2}{45} 
- \frac{F_3}{9} - \frac{F_4}{9} \right) .
\label{Udefinition}
\end{equation}
$U$ defines an average Coulomb repulsion within the $n$-holes multiplets, so
that
\begin{equation}
E^{\rm ave}(n)=
{\rm Tr}_n ( \hat{H}_0 + \hat{H}_{\rm  e-e} )=
\epsilon  \, n + U \frac {n (n-1)}2 \, .
\end{equation}
 It should be noted that $U$ differs from the usual definition of the
 Hubbard $U$, involving the lowest multiplet in each $n$-configuration: $
 U^{\rm min} = E^{\rm min}(n+1) + E^{\rm min}(n-1) - 2 E^{\rm min}(n)
 $. This second definition is inconvenient here, since it depends wildly
 on $n$. 
%

\section{The adiabatic calculations}
\label{adiabatic:sec}

We approximate the vibron operators $\hat{Q}_{i\Lambda \mu}$ with classical
coordinates, in the spirit of the adiabatic approximation.  
In an orbitally degenerate situation (as for the C$_{60}^{n+}$ ions at
hand) the adiabatic approximation usually yields fairly accurate energetics
in the limit of large the e-v couplings, so that tunneling between
equivalent minima can safely be neglected \cite{Englman}.
The phonon kinetic term in (\ref{vib-hamiltonian}) is neglected in the
adiabatic approximation. In Sect.~\ref{nonadiabatic:sec} we will partly
restore this term by taking into account quantum zero-point energies.
In any classical statically JT-distorted configuration,
the icosahedral symmetry is broken:
therefore states of different icosahedral symmetry representations are
inter-mixed.
Only the total number of holes $n$, total spin $S$ and its projection $S_z$
are conserved upon distortion.
Here, we neglect any change of the Coulomb Hamiltonian upon distortion, and
we assume therefore that $\hat{H}_{\rm e-e}$ is still determined according
to Eqs.\
(\ref{Coulomb-hamiltonian},\ref{Coulomb-ints},\ref{Fdecomposition}) by the
same parameters $F_i$ of Table~\ref{e-e-params:table}, as in icosahedral
symmetry.
Also, we assume no change of the phonon frequencies $\omega_{i\Lambda}$ and
couplings $g^r_{i\Lambda}$ upon charging.

\begin{table}[tb]
\begin{center}
\begin{tabular}{cc|rrr}
\hline
\hline
$n$&$S$ & adiabatic & vibrational & electronic \\
\hline
2   &   0   &    -129    &   270   &    -399 \\
    &   1   &{\bf-142}   &   99    &    -241 \\
& & \\
3   &   1/2  &    -168    &   267   &    -435\\
    &   3/2  &{\bf-222}   &   99    &    -320\\
& & \\
4   &   0    &    -200    &   361   &    -561\\
    &   1    &    -211    &   229   &    -440\\
    &   2    &{\bf-308}   &   69    &    -377\\
& & \\
5   &   1/2  &    -203    &   308   &    -511\\	
    &   3/2  &    -256    &   169   &    -425\\
    &   5/2  &{\bf-397}   &   0     &    -397\\
\hline
\hline
\end{tabular}
\end{center}
\caption{
The total adiabatic energy $V^{\rm ad}({\bf Q}_{\rm min})$ (in meV) of the
lowest electronic state for each $n$ and $S$, including the e-e and e-v
contributions from $\hat{H}_{\rm vib} +\hat{H}_{\rm e-v} +\hat{H}_{\rm
e-e}$ (but excluding the $\left[U n(n-1)/2\right]$ term), for
C$_{60}^{n+}$.
The last two columns distinguish the vibrational ($\hat{H}_{\rm vib}$) and
electronic ($\hat{H}_{\rm e-v} + \hat{H}_{\rm e-e}$) contributions.
\label{adiabatic_energies}
}
\end{table}

For each $n$, $S$ and $M_S$, we allow the 64 ($6\times 4$ $G_g$ plus
$8\times 5$ $H_g$) phonon coordinates to relax, and determine the optimal
distortion, by full minimization of the lowest adiabatic potential sheet
$V^{\rm ad}({\bf Q})$
in the space of all the phonons coordinates ${\bf Q}$.
We leave the $A_g$ modes out, since they contribute a trivial
\begin{equation}
E^{A_g}(n)= -\frac 18 n^2 \sum_i g_{i A_g}^2 \hbar \omega_{i A_g} =
            -n^2\cdot 1.79~{\rm meV},
\label{ag_energy}
\end{equation}
spin- and symmetry-independent term to the energetics.
Because of particle-hole symmetry, charges $n>5$ can always be reduced to
the computed charges $n<5$.
In Table~\ref{adiabatic_energies}, we report the resulting
optimally-distorted energy in each spin sector, based on the
electron-electron (e-e) and e-v couplings of C$_{60}^{n+}$ ions, as
previously published in Ref.~\cite{Lueders02}.
The main outcome of the adiabatic calculation is that positive C$_{60}$
ions favor high-spin ground states (contrary to the analogous finding for
negative ions).

In the present contribution, we extend the previous calculation to obtain
the complete set of {\em all} the equivalent minima for each $(n,S)$
sector.
To this purpose, we generate about a hundred randomly distributed
distortions away from the $I_h$ high-symmetry point, and let the
molecule relax to the closest minimum, by combined standard (simplex and
conjugate-gradients) minimization algorithms.
We then apply the symmetry operations of the icosahedral group to the each
of the minima found, in order to locate any possibly missing minimum.
Although the method employed is not deterministic, the symmetry analysis
makes the probability that any set of minima is incomplete utterly
negligible.
Thus, for each $n$ and $S$ we obtain a complete set of equivalent global
minima.
In the few cases where the minimization leads to non-global minima, we have
discarded them based on simple comparison of the adiabatic energies.

\begin{table}[tb]
\begin{center}
\begin{tabular}{cc|ccll}
\hline
\hline
$n$     &$S$    &       number of &     local   &       number of       &       distortion      \\
        &       &       minima    &     symmetry&       1$^{\rm st}$, 2$^{\rm nd}$, 3$^{\rm rd}$... neighbors   &       $\left|{\bf Q}_{\rm min}\right|$        \\
\hline
2       &       0       &       6       &$D_{5d}$       &       5                       &       3.12\\
        &       1       &       15      &$D_{2h}$       &       4  4  4  2      &       1.87\\
& \\
3       &       1/2     &       30      &$C_{2v}$       &       2 1 2 4 4 2 2 2 4 4 2   &       3.08\\
        &       3/2     &       15      &$D_{2h}$       &       4  4  4  2              &       1.87\\
&\\
4       &       0       &       10      &$D_{3d}$       &       3  6            &       3.52\\
        &       1       &       30      &$C_{2v}$       &       2 2 2 2 1 4 4 4 6 2     &       2.85\\
        &       2       &       6       &$D_{5d}$       &       5                       &       1.58\\
& \\
5       &       1/2     &       60      &$C_{2v}$       &       1 2 2 4 4 4 2 2 2 4 2  &       3.27\\
        &               &  &  &        \quad 2 4 2 2 2 4 4 4 2 2 1 1  & \\
        &       3/2     &       30      &$C_{2v}$       &       8  12 8  1              &       2.46\\
        &       5/2     &       1       &$I_h$          &       0               &       0\\
\hline
\hline
\end{tabular}
\caption{
\label{tab-symm}
The number and the local symmetries of the JT minima for given
charge $n$ and spin $S$.
In the 5th column the number of neighbors of all orders are listed for
a given minimum.
The last column gives the total amount of dimensionless JT distortion at
each minimum.
}
\end{center}
\end{table}

In Table \ref{tab-symm} we summarize some global properties of the obtained
JT minima for all charge and spin states.
In these multi-mode JT systems, the local symmetry of an optimal
distortion is described in terms of the subgroup $G_{\rm local}$ of
symmetry operations which leave that minimum invariant.
We remind that the minima in the $n=1$ $S=\frac 12$ case, where e-e
interaction is unimportant, were found to be 6, of local $D_{5d}$ symmetry
\cite{CeulemansII,hbyh,Manini01}.
For $2\leq n\leq 8$, where the role of e-e interaction is crucial, the
number of JT minima follows from the local symmetry: it is generally given
by the ratio $|I_h|/|G_{\rm local}|$ of the orders of the icosahedral group
(120) and of the invariant subgroup.

Special care has to be taken for $n=5$ holes.
Here, in addition to the icosahedral symmetry, the system is particle-hole
symmetric, i.e.\ invariant under exchange 
of fermion creation and
annihilation operators. 
This transformation leaves the Coulomb Hamiltonian $\hat{H}_{\rm e-e}$
invariant, while the the vibron interaction $\hat{H}_{\rm e-v}$ is
unchanged provided that a sign change of the vibron coordinates
$\hat{Q}_{i,\Lambda,\mu}$ is also performed.
Hence, for a given minimum ${\bf Q}_{\rm min}$, also its opposite $-{\bf
Q}_{\rm min}$ is a minimum of the potential energy surface.
In the case $n=5$ $S=\frac 12$, this leads to a doubling of the 
minima: the local $C_{2v}$ symmetry would lead to 30 minima,
but 30 more equivalent minima are added in the opposite positions by
particle-hole symmetry.
For $n=5$ $S=\frac 32$ instead, the number of minima remains
30, since for each minimum there is one of the $I_h$ symmetry operations
that transforms this minimum into its opposite point.
Note that this operation is not the spatial inversion (since all vibrations
considered here form even representations), which are invariant under
inversion, but a twofold rotation.
Finally, for $n=5$ $S=\frac 52$, the electronic state is orbitally
nondegenerate, thus no JT distortion takes place.

\begin{table}
\begin{center}
\begin{tabular}{cc|cl}
\hline
\hline
$n$     &$S$    & Symmetry &      distortions of $G_g$ and $H_g$ modes (dimensionless)\\
\hline
2       &       0       & $D_{5d}$ &      0.00 0.00 0.00 0.00 0.00 0.00\\
        &               &        &      \quad 2.69 0.935 0.0095 0.692 0.0455 0.405 0.749 0.657 \\
2       &       1       & $D_{2h}$ &      0.0755 0.0102 0.0799 0.0623 0.0227 0.0466 \\
        &               &  &     \quad 1.58 0.611 0.138 0.404 0.0516 0.262 0.486 0.346   \\
& & & \\
3       &       1/2     & $C_{2v}$ &      0.0548 0.0074 0.0580 0.0452 0.0165 0.0339\\
        &               &  &     \quad 2.62 1.01 0.185 0.669 0.0799 0.433 0.801 0.571    \\
3       &       3/2     & $D_{2h}$ &     0.0755 0.0102 0.0799 0.0623 0.0227 0.0466\\
        &               & &       \quad 1.58 0.611 0.138 0.404 0.0516 0.262 0.486 0.346   \\
& & & \\
4       &       0       & $D_{3d}$ &      0.0828 0.0112 0.0877 0.0683 0.0249 0.0512\\
        &               & &      \quad 2.88 1.30 0.494 0.728 0.153 0.553 1.02 0.486       \\
4       &       1       & $C_{2v}$ &     0.074 0.010 0.0782 0.061 0.0223 0.0457\\
        &               & &      \quad 2.39 0.968 0.228 0.609 0.0879 0.414 0.767 0.486   \\
4       &       2       & $D_{5d}$ &      0.00 0.00 0.00 0.00 0.00 0.00\\
        &               & &      \quad 1.36 0.473 0.0046 0.35 0.023 0.205 0.379 0.333    \\
& & & \\
5       &       1/2     & $C_{2v}$ &      0.101 0.0135 0.106 0.0827 0.0302 0.0619\\
        &               & &      \quad 2.70 1.17 0.401 0.683 0.129 0.50 0.926 0.492        \\
5       &       3/2     & $C_{2v}$ &      0.0384 0.0053 0.0411 0.032 0.0117 0.024\\
        &               & &      \quad 2.12 0.756 0.0391 0.544 0.0427 0.327 0.605 0.503  \\
5       &       5/2     & $I_h$ &      0.00 0.00 0.00 0.00 0.00 0.00\\
        &               & &      \quad 0.00 0.00 0.00 0.00 0.00 0.00 0.00 0.00\\
\hline
\end{tabular}
\end{center}
\caption{
The JT distortion at the minima, for each mode and value of the charge $n$
and spin $S$.
The distortions $|\hat{Q}_{i\Lambda}|$ are given in units of the length
scale $x_0(\omega)=\sqrt{\hbar/ (\omega \, m_{\rm C})}$ associated to each
harmonic oscillator ($m_{\rm C}$ is the mass of the C atom).
The $x_0(\omega_{i\Lambda})$ for the $G_g$ and $H_g$ modes of
C$_{60}$ are: 76.3, 70.4, 60.3, 50.3, 46.1, 43.0; 103.7, 80.9, 62.6, 59.8,
50.1, 47.0, 43.9, 42.1 pm, respectively.
\label{tab-mode}}
\end{table}

Table~\ref{tab-mode} collects some quantitative information about the
contribution of each mode to the amount of JT distortion at each minimum.
As expected, the largest distortion involves always
the lowest $H_g$ mode, which is the most strongly coupled one (see
Table~\ref{e-v-params:table}).
Also, the $D_5$ distortions receive no contribution of the $G_g$ modes,
which contribute to all the lower-symmetry minima instead.

Table \ref{tab-symm} contains some information about the connectivity of
the minima in ${\bf Q}$ space.
In some cases, the specification of the number of first, second,
etc.\ neighbors of a given minimum is sufficient to clarify completely the
topology of the minima in the 64-dimensional space.
In particular, the $D_{5d}$ wells of both the $n=2$ $S=0$ and the $n=4$ $S=2$
surfaces are located on the six vertices of the five-dimensional
regular simplex, the generalization of a tetrahedron, each minimum being
equally distant from all the others: this is analogous to the previously
determined minima of $V^{\rm ad}$ for $n=1$ $S=\frac 12$ \cite{hbyh}.
In analogy, the connectivity of the 10 $D_{3d}$ minima for $n=4$ $S=0$ is
the same of the one depicted in Fig.~1b of Ref.~\cite{hbyh}.

For the other cases of lower symmetry, the number of neighbors of any given
order must be complemented by some extra connectivity information.
First, we observe that the minima for $n=2$ $S=1$ and for $n=3$ $S=\frac
32$ are exactly the same.
Indeed, these two case are related by a particle-hole symmetry applied only
to one spin flavor.
For all nonequivalent cases, the complete topological information about the
wells is contained in the connectivity matrix ${\cal C}\left(n,S\right)$,
whose matrix elements indicate that minima $i$ and $j$ are ${\cal
C}\left(n,S\right)_{ij}$th neighbors.
We report those matrices in the Appendix.
Careful exam of ${\cal C}\left(n,S\right)$ for $n=2$ $S=1$ and $n=3$
$S=\frac 32$ shows that each of the 15 minima is linked to four nearest
neighbor minima, which, in turn, are linked to more minima, forming a
completely connected regular polytope.
%
The matrices ${\cal C}\left(3,\frac 12\right)$ and ${\cal
C}\left(4,1\right)$, show that, for $n=3$ $S=\frac 12$ and $n=4$ $S=1$,
the 30 minima are divided into 6 pentagonal ``clusters'' of five
nearest-neighboring minima.
In contrast, for $n=5$ $S=\frac 12$, nearest-neighbor wells come in pairs.
Finally, the 30 minima for $n=5$ $S=\frac 32$, show the largest
connectivity, and sit at the vertices of a highly symmetric polytope.

\section{Vertical excitation energies}
\label{excitations:sec}

\begin{table}
\begin{center}
\begin{tabular}{cc|c|cc}
\hline
\hline
\multicolumn{2}{c|}{minimum}    &exc.\ states&  $\Delta E^{\rm min}$    &       $\Delta E^{\rm max}$    \\
$n$     &       $S$     &       $S'$    &       [meV]   &       [meV]   \\
\hline
2       &       0       &       0       &       221     &       823     \\
        &               &       1       &       140     &       507     \\
	&       1       &       0       &       75      &       635     \\
        &               &       1       &       127     &       346     \\
3       &       1/2     &       1/2     &       132     &       918     \\
        &               &       3/2     &       99      &       619     \\
	&       3/2     &       1/2     &       125     &       784     \\
        &               &       3/2     &       127     &       346     \\
4       &       0       &       0       &       192     &       1464    \\
        &               &       1       &       102     &       1141    \\
        &               &       2       &       165     &       509     \\
	&       1      &       0       &       82      &       1232    \\
        &               &       1       &       120     &       881     \\
        &               &       2       &       54      &       408     \\
4       &       2       &       0       &       234     &       981     \\
        &               &       1       &       163     &       647     \\
        &               &       2       &       167     &       179     \\
5	&       1/2    &       1/2     &       128     &       1320    \\
        &               &       3/2     &       74      &       805     \\
        &               &       5/2     &       114     &       -       \\
	&       3/2     &       1/2     &       120     &       997     \\
        &               &       3/2     &       143     &       693     \\
        &               &       5/2     &       28      &       -       \\
	&       5/2     &       1/2     &       316     &       731     \\
        &               &       3/2     &       203     &       377     \\
        &               &       5/2     &       -       &       -       \\
\hline
\end{tabular}
\end{center}
\caption{The lowest and highest vertical excitation energies (in meV) 
	calculated assuming that the C$_{60}^{n+}$ ion remains frozen in
        one of the adiabatic minima when the electronic state is excited.
        The first two columns fix the relevant distortion.  The third
        column indicates the spin $S'$ of the excited states considered.
        The excitation energies in the last two columns are referred to the
        adiabatic energy of each specific minimum, reported in
	Table~\protect\ref{adiabatic_energies}. 
\label{tab-excitationrange}}
\end{table}

In Table~\ref{tab-excitationrange} we report the range of ``vertical''
excitation energies $\Delta E$ for all final spin symmetries $S'$,
in the frozen
minimum configurations ${\bf Q}_{\rm min}(n,S)$, for all values of $n$ and
$S$.
The complete spectrum (available upon request from the authors) is very
dense and not much informative.
The listed energies give a quantitative prevision of the spectral range
where a fast (optical) spectroscopy is likely to locate the intra-band HOMO
excitations of the C$_{60}^{n+}$ ions.
For the experimentally most accessible case $n=2$ $S=1$, here follows the
complete list of the triplet-triplet excitation energies: 127, 149, 150,
178, 182, 218, 326, 337, and 346 meV.

\section{Non-adiabatic corrections}
\label{nonadiabatic:sec}

The leading quantum correction to the static JT energetics is given by the
zero-point energy gain due to the softening of the vibrational frequency at
the JT-distorted minima \cite{ManyModes}.
To obtain this information, by finite differences we compute the Hessian
matrix of the second-order derivatives of the lowest adiabatic potential
sheet, at one of the static JT minima ${\bf Q}_{\rm min}$
\begin{equation}
{\cal H}_{\{i\Lambda \mu\} \{i'\Lambda' \mu'\} }=
\left. \frac {\partial^2 V^{\rm ad}({\bf Q})}
{\partial Q_{i\Lambda \mu}\partial Q_{i'\Lambda' \mu'}}
\right|_{{\bf Q}_{\rm min}}
\end{equation}
The vibrational frequencies $\tilde\omega_j$ at the distorted point are
the eigenvalues of the Hessian matrix.
There is no square root involved in the $\tilde\omega_j$, as the
coordinates are scaled with the harmonic length scale $x_0(\omega)$ defined
in Table~\ref{tab-mode}.
In the harmonic approximation, these ``new'' normal mode frequencies
$\tilde\omega_j$ contribute a zero-point energy $\sum_j \frac 12
\hbar\tilde\omega_j$ to that minimum configuration: the difference between
this and the original zero-point energy $\sum_{i\Lambda \mu} \frac 12 \hbar
\omega_{i\Lambda}$ gives the leading quantum correction 
\begin{equation}
E_{\rm zero}(n,S) =
 \frac 12 \left[
\sum_j \hbar\tilde\omega_j (n,S)
-\sum_{i\Lambda \mu}  \hbar \omega_{i\Lambda}
\right] ,
\label{Ezer}
\end{equation}
to the classical energy.
We rewrite this correction as
\begin{eqnarray}
E_{\rm zero}(n,S) &=&
\frac 12 \sum_{i\Lambda \mu}
\left[
{\cal H}_{\{i\Lambda \mu\} \{i\Lambda \mu\} }(n,S)
- \hbar \omega_{i\Lambda} 
\right]
\nonumber\\
&=& \frac 12 \left[
\left. \nabla^2 V^{\rm ad}({\bf Q})\right|_{{\bf Q}_{\rm min}}
- \sum_{i\Lambda \mu} \hbar \omega_{i\Lambda}
\right],
\label{Ezero}
\end{eqnarray}
using the invariance of the trace under change of basis.

\begin{table}[tb]
\begin{center}
\begin{tabular}{cc|rrc}
\hline \hline
$n$ & $S$ & $E_{\rm class}$& $E_{\rm zero}$ & 
$E_{\rm class} + E_{\rm zero}$ \\
\hline
2   &   0   &    -129    &     -125    &    -254    \\
    &   1   &{\bf-142}   &     -159    &{\bf-301}   \\
& & \\
3   &   1/2  &    -168    &    -207    &    -376    \\
    &   3/2  &{\bf-222}   &    -159    &{\bf-380}   \\
& & \\
4   &   0    &    -200    &    -213    &    -412    \\
    &   1    &    -211    &    -227    &{\bf-437}   \\
    &   2    &{\bf-308}   &     -93     &    -400    \\
& & \\
5   &   1/2  &    -203    &    -247    &{\bf-449}   \\	
    &   3/2  &    -256    &    -175    &    -431    \\
    &   5/2  &{\bf-397}   &       0    &    -397    \\
\hline
\end{tabular}
\caption{
The total adiabatic energy $E_{\rm class}=V^{\rm ad}({\bf Q}_{\rm min})$
(in meV) of the lowest electronic state for each $n$ and $S$, including the
e-v and e-e contributions from $\hat{H}_{\rm vib} +\hat{H}_{\rm e-v}
+\hat{H}_{\rm e-e}$ (but excluding the $\left[U n(n-1)/2\right]$ term), for
C$_{60}^{n+}$.
The following column contains the leading non-adiabatic correction $E_{\rm
zero}$, the  zero-point energy defined in Eq.\ (\ref{Ezero}).
The last column reports the adiabatic energy $E_{\rm class}$ corrected by
the zero-point term $E_{\rm zero}$: for $n\geq 4$ it leads to a
different ordering of the spin states.
\label{energies:tab}
}
\end{center}
\end{table}

Table \ref{energies:tab} displays the lowest adiabatic energies for given
charge and spin in various approximations. The first column reports the
adiabatic energy, $E_{\rm class}=V^{\rm ad}({\bf Q}_{\rm min})$, as in
Table~\ref{adiabatic_energies}.
The zero-point energy corrections in the following column are comparable
in magnitude to the leading adiabatic energies.
In particular the very large values of zero-point energy gain for $n=5$
$S=\frac 12$ is associated to  very shallow minima, connected by low
barriers. 
The lowest vibrational frequency is as small as $\tilde\omega_1\approx 2.3$
meV.
In the rather close competition between the Coulomb physics (Hund's rules)
and the JT physics (anti-Hund behavior) the zero-point correction
is very important, and, as shown by the last column of
Table~\ref{energies:tab}, reduces drastically the spin-gap ($n=3$), or even
changes the ground-state symmetry in favor of an intermediate ($n=4$) or low
($n=5$) spin state.

The zero-point correction treated here represents the $g_i^0$ term of a
large-coupling expansion, where the adiabatic energy $E_{\rm class}$ is the
leading ($g_i^2$) term.
The next corrections to be considered, of order $g_i^{-2}$, are associated
to tunneling among minima, possibly affected by Berry phases
\cite{hbyh,AMT,noberry}.
Tunneling is likely to be especially important between the pairs of neighboring
minima of the $n=5$ $S=\frac 12$ adiabatic surface.
Tunneling will be dealt with in future work.

\rem{
Include also the effective $U^{\rm min}$ accounting for zero-point correction?
I don't know, it is probably not changing much from 3 eV.  Martin, try
compute it...
}


\section{Discussion and Conclusions}
\label{discussion:sec}

In the present calculation both e-e and e-v interactions are included for
the HOMO shell of C$_{60}$.
E-e exchange terms are treated essentially exactly, in the assumptions that
(i) inter-band couplings can be neglected, and only act as a renormalization of
the Coulomb parameters and that (ii) the latter are independent of the charge
$n$ in the HOMO.
In principle, due to both orbital and geometrical relaxation, the effective
Coulomb interaction (\ref{Coulomb-ints}) will depend on the instantaneous
charge state of the fullerene ion.
However, this effect, a very important one in single-atom calculations, is
expected to be fairly small in such a large molecule as C$_{60}$.
In a JT system, the coupled phonons should in principle be treated
fully quantum mechanically, as nonadiabatic effects may be important.
However for strong electron-phonon coupling, the leading terms
(of order $g^2$) are obtained in the adiabatic approximation, by studying
the minima of the lowest adiabatic potential surface $V^{\rm ad}({\bf Q})$.
Non-adiabatic effects are taken into account to the next order ($g^0$)
by the calculation of the new harmonic  oscillation frequencies close to
the adiabatic JT minima.
These zero-point corrections are significantly large,
and they can even reverse the
theoretical prevision for the symmetry of the ground state of the
C$_{6}^{n+}$ ion for $4\leq n\leq 6$.
Tunneling matrix elements which mix different minima to suitable dynamical
combinations restore the original icosahedral symmetry and provide the
next-order ($g^{-2}$) quantum correction to the energetics.
These terms, which will be the subject of future work, are likely to be
especially large for $n=5$ $S=\frac 12$.

The present calculation was carried out in the {\em linear} e-v
approximation.
As the coupling and thus the distortions are fairly large, quadratic and
higher-order (in ${\bf Q}$) couplings and vibrations anharmonicity could be
important.
Unfortunately, no estimate for those higher-order couplings is available yet.

The parameters used in this calculation, both for e-e and e-v interaction
are most likely underestimated by the local density approximation used in
their determination, as discussed in Ref.~\cite{Lueders02,Manini01}.
Consequently, both the Coulomb repulsion and the phonon-mediated attraction
calculated within the local density approximation are likely to need a
rescaling by a similar factor of order two.
Indeed, the balance between the two opposing interactions is delicate in
C$_{60}^{n-}$ ions (as demonstrated by the presence of both high-spin and
low-spin local ground states in different chemical environments
\cite{Brouet01,Kiefl,Zimmer95,Lukyanchuk95,Prassides99,Schilder94,Arovas95}).
In C$_{60}^{n+}$ e-e interaction prevails at the adiabatic level: high-spin
states are favored, as experiments confirm for $n=2$ \cite{Panich02}.
According to our calculation, however, more highly charged states, close to
HOMO half filling should favor local low-spin.

An effective local exchange interaction favoring low spin is a crucial
ingredient for superconductivity in a strongly-correlated orbitally-degenerate
material such as a solid of doped C$_{60}$ \cite{Capone01,Capone02}.
If screening and retardation effects could be neglected, the present
single-molecule calculation suggests that superconductivity should be
strongly suppressed in the hole-doped solid at doping $0\leq n\leq 3$, but
could be recovered close to half filling $4<n < 6$.
It is presently unclear if such a high level of hole doping is practically
accessible, except possibly by field-induced charging \cite{Schon00}.

\appendix
\section{Appendix}

We report here the connectivity matrices whose matrix element ($ij$)
indicates that minima $i$ and $j$ are ${\cal C}\left(n,S\right)_{ij}$th
neighbors.  for all the nontrivial $n$, $S$ cases.
We include up to 9th neighbors, substituting those of higher order with a
dash.

\rem{
\begin{equation} {\cal C}\left(2,0\right)=
\left[\matrix
{ - & 1 & 1 & 1 & 1 & 1 \cr  & - & 1 & 1 & 1 & 1 \cr  &  & - & 1 & 1 & 1 \cr 
 &  &  & - & 1 & 1 \cr  &  &  &  & - & 1 \cr  &  &  &  &  & - \cr  } 
\right]
\end{equation}
}

\begin{equation} {\cal C}\left(2,1\right)= {\cal C}\left(3,\frac 32 \right)=
 \quad\quad\quad\quad\quad\quad \quad\quad\quad\quad\quad\quad\quad
 \quad\quad\quad\quad\quad\quad \quad\quad
\label{C2-1}
\end{equation}
$$
\\
\left[\matrix
{ - & 4 & 4 & 2 & 3 & 1 & 3 & 2 & 1 & 1 & 2 & 3 & 3 & 1 & 2 \cr  & - & 4 & 
3 & 1 & 2 & 1 & 3 & 2 & 2 & 3 & 1 & 1 & 2 & 3 \cr  &  & - & 1 & 2 & 3 & 2 & 
1 & 3 & 3 & 1 & 2 & 2 & 3 & 1 \cr  &  &  & - & 4 & 4 & 3 & 2 & 1 & 2 & 3 & 
1 & 2 & 3 & 1 \cr  &  &  &  & - & 4 & 1 & 3 & 2 & 3 & 1 & 2 & 3 & 1 & 2 \cr  
&  &  &  &  & - & 2 & 1 & 3 & 1 & 2 & 3 & 1 & 2 & 3 \cr  &  &  &  &  &  & 
- & 4 & 4 & 1 & 2 & 3 & 2 & 3 & 1 \cr  &  &  &  &  &  &  & - & 4 & 3 & 1 & 
2 & 1 & 2 & 3 \cr  &  &  &  &  &  &  &  & - & 2 & 3 & 1 & 3 & 1 & 2 \cr  &  
&  &  &  &  &  &  &  & - & 4 & 4 & 2 & 3 & 1 \cr  &  &  &  &  &  &  &  &  &  
& - & 4 & 3 & 1 & 2 \cr  &  &  &  &  &  &  &  &  &  &  & - & 1 & 2 & 3 \cr  
&  &  &  &  &  &  &  &  &  &  &  & - & 4 & 4 \cr  &  &  &  &  &  &  &  &  &  
&  &  &  & - & 4 \cr  &  &  &  &  &  &  &  &  &  &  &  &  &  & - \cr  }
\right] 
$$

\begin{equation} {\cal C}\left(3,\frac 12\right)=
 \quad\quad\quad\quad\quad\quad \quad\quad\quad\quad\quad\quad\quad
 \quad\quad\quad\quad\quad\quad \quad\quad\quad\quad\quad\quad\quad
\end{equation}
$$
{\tiny
\left[\matrix
{ - & - & - & 2 & - & 6 & 3 & 5 & 9 & - & 3 & 5 & 9 & 5 & 8 & 9 & 7 & 1 & 
5 & 4 & 8 & 9 & 1 & 7 & 4 & 4 & 6 & - & - & 4 \cr  & - & - & - & 4 & 5 & 9 
& 8 & 1 & 6 & - & 9 & 4 & - & 4 & 6 & 9 & 5 & 9 & 5 & 1 & 4 & 7 & 8 & 3 & 
5 & 3 & - & 2 & 7 \cr  &  & - & - & 5 & 9 & 4 & 1 & 7 & 3 & 6 & 4 & 5 & 4 & 
7 & 5 & 1 & 8 & 6 & - & 5 & 3 & 9 & 4 & 9 & 9 & - & 2 & - & 8 \cr  &  &  & 
- & 9 & 4 & 5 & 7 & 8 & 9 & 5 & 7 & 8 & 3 & 9 & - & 5 & 4 & 3 & 6 & 9 & - & 
4 & 5 & 6 & 1 & 4 & - & - & 1 \cr  &  &  &  & - & - & - & - & 2 & - & 5 & 
3 & 9 & 8 & 4 & 1 & 9 & 7 & 4 & 8 & 5 & 7 & 9 & 1 & 5 & 6 & 4 & 3 & 6 & 
- \cr  &  &  &  &  & - & - & - & - & 4 & 8 & 9 & 1 & 4 & 3 & 5 & 6 & 9 & 5 
& 1 & 9 & 8 & 4 & 7 & - & 5 & 7 & - & 3 & 2 \cr  &  &  &  &  &  & - & 2 & 
- & 5 & 1 & 4 & 7 & 7 & 9 & 8 & 5 & 1 & - & 5 & 6 & 4 & 3 & 9 & 4 & 9 & 8 & 
6 & - & - \cr  &  &  &  &  &  &  & - & - & 3 & 4 & 6 & 5 & 5 & - & 9 & 3 & 
4 & 9 & 7 & 4 & 1 & 5 & 8 & 6 & - & 9 & 4 & 9 & - \cr  &  &  &  &  &  &  &  
& - & 9 & 7 & 5 & 8 & 9 & 6 & 4 & - & 5 & 6 & 9 & 3 & 5 & - & 4 & 3 & 4 & 
1 & 5 & 4 & - \cr  &  &  &  &  &  &  &  &  & - & - & - & 2 & 4 & 5 & 7 & 1 
& 9 & 8 & 5 & 4 & 1 & 7 & 9 & 8 & - & - & 5 & 4 & 6 \cr  &  &  &  &  &  &  & 
 &  &  & - & 2 & - & - & 6 & 4 & 9 & 3 & 7 & 4 & 9 & 8 & 1 & 5 & 5 & - & - 
& 4 & 9 & 9 \cr  &  &  &  &  &  &  &  &  &  &  & - & - & 9 & 4 & 1 & 8 & 5 
& 5 & 6 & - & 9 & 4 & 3 & 7 & - & - & 1 & 8 & - \cr  &  &  &  &  &  &  &  &  
&  &  &  & - & 6 & 3 & 5 & 4 & - & 9 & 3 & 6 & 4 & 5 & - & 9 & - & - & 7 & 
1 & 4 \cr  &  &  &  &  &  &  &  &  &  &  &  &  & - & - & - & 2 & - & 1 & 8 
& 7 & 5 & 9 & 4 & - & 3 & 5 & 6 & 9 & 1 \cr  &  &  &  &  &  &  &  &  &  &  & 
 &  &  & - & 2 & - & - & 7 & 1 & 8 & 9 & 4 & 5 & - & - & 9 & 5 & 1 & 5 \cr  
&  &  &  &  &  &  &  &  &  &  &  &  &  &  & - & - & - & 5 & 4 & 9 & - & 6 & 
3 & - & 9 & 8 & 3 & 4 & 7 \cr  &  &  &  &  &  &  &  &  &  &  &  &  &  &  &  
& - & - & 4 & 9 & 5 & 3 & - & 6 & - & 5 & 7 & 4 & 8 & 4 \cr  &  &  &  &  &  
&  &  &  &  &  &  &  &  &  &  &  & - & 9 & 5 & 4 & 6 & 3 & - & 2 & 6 & 4 & 
9 & 7 & 8 \cr  &  &  &  &  &  &  &  &  &  &  &  &  &  &  &  &  &  & - & - & 
- & - & - & 2 & 8 & 1 & 4 & 4 & - & 3 \cr  &  &  &  &  &  &  &  &  &  &  &  
&  &  &  &  &  &  &  & - & - & - & 2 & - & 7 & 9 & - & 9 & 3 & 4 \cr  &  &  
&  &  &  &  &  &  &  &  &  &  &  &  &  &  &  &  &  & - & 2 & - & - & 1 & 5 
& 3 & 7 & 4 & - \cr  &  &  &  &  &  &  &  &  &  &  &  &  &  &  &  &  &  &  & 
 &  & - & - & - & 4 & 7 & 5 & 5 & 6 & 9 \cr  &  &  &  &  &  &  &  &  &  &  & 
 &  &  &  &  &  &  &  &  &  &  & - & - & 5 & 8 & 9 & 8 & 5 & 6 \cr  &  &  &  
&  &  &  &  &  &  &  &  &  &  &  &  &  &  &  &  &  &  &  & - & 9 & 4 & 6 & 
1 & 9 & 5 \cr  &  &  &  &  &  &  &  &  &  &  &  &  &  &  &  &  &  &  &  &  & 
 &  &  & - & 4 & 1 & - & 5 & 9 \cr  &  &  &  &  &  &  &  &  &  &  &  &  &  & 
 &  &  &  &  &  &  &  &  &  &  & - & 2 & 8 & 7 & 3 \cr  &  &  &  &  &  &  &  
&  &  &  &  &  &  &  &  &  &  &  &  &  &  &  &  &  &  & - & 9 & 5 & 5 \cr  & 
 &  &  &  &  &  &  &  &  &  &  &  &  &  &  &  &  &  &  &  &  &  &  &  &  &  
& - & - & 9 \cr  &  &  &  &  &  &  &  &  &  &  &  &  &  &  &  &  &  &  &  &  
&  &  &  &  &  &  &  & - & 5 \cr  &  &  &  &  &  &  &  &  &  &  &  &  &  &  
&  &  &  &  &  &  &  &  &  &  &  &  &  &  & - \cr  } 
\right]}
$$

\rem{
\begin{equation} {\cal C}\left(4,0\right)=
\pmatrix
{ - & 2 & 2 & 2 & 2 & 2 & 2 & 1 & 1 & 1 \cr  & - & 2 & 2 & 2 & 1 & 1 & 2 & 
2 & 1 \cr  &  & - & 1 & 1 & 2 & 2 & 2 & 2 & 1 \cr  &  &  & - & 2 & 2 & 1 & 
2 & 1 & 2 \cr  &  &  &  & - & 1 & 2 & 1 & 2 & 2 \cr  &  &  &  &  & - & 2 & 
2 & 1 & 2 \cr  &  &  &  &  &  & - & 1 & 2 & 2 \cr  &  &  &  &  &  &  & - & 
2 & 2 \cr  &  &  &  &  &  &  &  & - & 2 \cr  &  &  &  &  &  &  &  &  & - 
\cr  } 
\end{equation}
}

\begin{equation} {\cal C}\left(4,1\right)= 
 \quad\quad\quad\quad\quad\quad \quad\quad\quad\quad\quad\quad\quad\quad
 \quad\quad\quad\quad\quad\quad \quad\quad\quad\quad\quad\quad\quad
\label{C4-1}
\end{equation}
$$
{\tiny
\left[\matrix
{ - & - & 8 & - & 2 & 8 & 9 & 5 & 6 & 2 & 4 & 6 & 7 & 9 & 9 & 6 & 7 & 7 & 
7 & 3 & - & 1 & 1 & - & 9 & 3 & 6 & 8 & 4 & 8 \cr  & - & 4 & - & 8 & 9 & 8 
& 7 & 8 & 9 & 6 & 4 & 5 & 2 & 3 & 3 & - & 1 & 7 & 9 & 7 & 7 & - & 1 & 6 & 
6 & 9 & 2 & 8 & 6 \cr  &  & - & 6 & - & 8 & 7 & 8 & 9 & 7 & 9 & - & 6 & 3 & 
7 & 2 & 4 & 6 & 1 & 6 & 3 & 9 & - & 7 & - & 1 & 9 & 8 & 2 & 5 \cr  &  &  & 
- & 9 & 2 & 2 & 7 & 4 & 8 & 8 & 8 & 7 & 8 & 6 & 9 & 1 & - & 5 & 6 & 1 & - & 
7 & 7 & 3 & 9 & 3 & 9 & 6 & 4 \cr  &  &  &  & - & 6 & 5 & 9 & 8 & 1 & 3 & 
7 & 8 & 6 & - & 9 & 6 & 3 & - & 4 & 9 & 2 & 1 & 7 & - & 7 & 6 & 4 & 8 & 
7 \cr  &  &  &  &  & - & 1 & 8 & 3 & 4 & - & 7 & - & 6 & 6 & - & - & 9 & 9 
& 9 & 1 & 7 & 3 & 6 & 4 & - & 7 & 5 & 7 & 8 \cr  &  &  &  &  &  & - & - & 
8 & 6 & 7 & - & 8 & 4 & 9 & - & 1 & 7 & 9 & 6 & 2 & 9 & 6 & 3 & 7 & - & 4 & 
6 & 8 & 3 \cr  &  &  &  &  &  &  & - & 1 & 9 & 6 & 1 & 3 & - & 2 & 4 & - & 
- & 3 & 7 & 9 & 6 & 6 & 9 & 2 & 7 & 4 & 8 & 6 & 8 \cr  &  &  &  &  &  &  &  
& - & 7 & 9 & 2 & 8 & - & 1 & 6 & - & - & 6 & - & 6 & 4 & 3 & 9 & 2 & 9 & 
7 & 7 & 5 & - \cr  &  &  &  &  &  &  &  &  & - & 8 & 8 & - & 5 & - & 6 & 3 
& 6 & 8 & 7 & 7 & 1 & 2 & 9 & - & 4 & 9 & 6 & 3 & - \cr  &  &  &  &  &  &  & 
 &  &  & - & 5 & 1 & 8 & 9 & - & - & 3 & 8 & 2 & - & 7 & 6 & 4 & 6 & 7 & 1 
& 7 & - & 2 \cr  &  &  &  &  &  &  &  &  &  &  & - & 6 & 8 & 2 & 7 & - & 7 
& 8 & 9 & 9 & 3 & 4 & 6 & 1 & - & 6 & 3 & 9 & 9 \cr  &  &  &  &  &  &  &  &  
&  &  &  & - & 9 & 7 & 7 & - & 6 & 3 & 2 & - & - & 9 & 6 & 4 & 4 & 2 & 9 & 
8 & 1 \cr  &  &  &  &  &  &  &  &  &  &  &  &  & - & 7 & 4 & 7 & 1 & 8 & - 
& 3 & 6 & 9 & 2 & 9 & 6 & - & 1 & 7 & 7 \cr  &  &  &  &  &  &  &  &  &  &  & 
 &  &  & - & 3 & 9 & 8 & 4 & - & 5 & 7 & 8 & 6 & 1 & 8 & 8 & 4 & 6 & - \cr  
&  &  &  &  &  &  &  &  &  &  &  &  &  &  & - & 8 & 6 & 2 & 8 & 7 & 5 & 9 & 
8 & 8 & 1 & - & 7 & 1 & 9 \cr  &  &  &  &  &  &  &  &  &  &  &  &  &  &  &  
& - & - & 6 & - & - & 8 & 4 & 8 & 8 & 7 & - & 9 & 3 & - \cr  &  &  &  &  &  
&  &  &  &  &  &  &  &  &  &  &  & - & 9 & 7 & 8 & 4 & 8 & 2 & 9 & 5 & 8 & 
2 & 9 & 4 \cr  &  &  &  &  &  &  &  &  &  &  &  &  &  &  &  &  &  & - & 4 & 
6 & 9 & - & - & 7 & 2 & 7 & - & 1 & 6 \cr  &  &  &  &  &  &  &  &  &  &  &  
&  &  &  &  &  &  &  & - & 9 & 8 & 6 & 8 & 8 & 3 & 1 & - & 7 & 1 \cr  &  &  
&  &  &  &  &  &  &  &  &  &  &  &  &  &  &  &  &  & - & - & 8 & 4 & 6 & 8 
& 8 & 6 & 4 & 7 \cr  &  &  &  &  &  &  &  &  &  &  &  &  &  &  &  &  &  &  & 
 &  & - & 2 & 8 & 8 & 6 & 9 & 3 & 6 & - \cr  &  &  &  &  &  &  &  &  &  &  & 
 &  &  &  &  &  &  &  &  &  &  & - & - & 7 & 8 & 5 & 7 & 7 & 9 \cr  &  &  &  
&  &  &  &  &  &  &  &  &  &  &  &  &  &  &  &  &  &  &  & - & 5 & 9 & 7 & 
1 & - & 3 \cr  &  &  &  &  &  &  &  &  &  &  &  &  &  &  &  &  &  &  &  &  & 
 &  &  & - & - & 3 & 6 & 9 & 7 \cr  &  &  &  &  &  &  &  &  &  &  &  &  &  & 
 &  &  &  &  &  &  &  &  &  &  & - & 8 & 9 & 2 & 6 \cr  &  &  &  &  &  &  &  
&  &  &  &  &  &  &  &  &  &  &  &  &  &  &  &  &  &  & - & - & - & 2 \cr  & 
 &  &  &  &  &  &  &  &  &  &  &  &  &  &  &  &  &  &  &  &  &  &  &  &  &  
& - & - & 8 \cr  &  &  &  &  &  &  &  &  &  &  &  &  &  &  &  &  &  &  &  &  
&  &  &  &  &  &  &  & - & 9 \cr  &  &  &  &  &  &  &  &  &  &  &  &  &  &  
&  &  &  &  &  &  &  &  &  &  &  &  &  &  & - \cr  }
\right].
}
$$

\rem{
\begin{equation} {\cal C}\left(4,2\right)=
\left[\matrix
{ - & 1 & 1 & 1 & 1 & 1 \cr  & - & 1 & 1 & 1 & 1 \cr  &  & - & 1 & 1 & 1 \cr 
 &  &  & - & 1 & 1 \cr  &  &  &  & - & 1 \cr  &  &  &  &  & - \cr  } \right].
\end{equation}
}

The 60 minima for $n=5$ $S=\frac 12$ are conveniently split into two sets
of 30 minima connected by the $I_h$ operations.
The minima in the ``b'' block are one by one ordinately opposite to those
in the ``a'' block.
Accordingly, the structure of the ${\cal C}$-matrix is as follows:
\begin{equation}
{\cal C}\left(5,\frac 12\right)=
\left[\matrix
{ {\cal C}\left(5,\frac 12\right)^{\rm a} &
        {\cal C}\left(5,\frac 12\right)^{\rm b} \cr
  {\cal C}\left(5,\frac 12\right)^{\rm b} &
        {\cal C}\left(5,\frac 12\right)^{\rm a} \cr
}\right]\ ,
\end{equation}
where 
\newpage
\begin{equation} {\cal C}\left(5,\frac 12\right)^{\rm a}=
 \quad\quad\quad\quad\quad\quad \quad\quad\quad\quad\quad\quad\quad
 \quad\quad\quad\quad\quad\quad \quad\quad\quad\quad\quad\quad
\end{equation}
$$
{\tiny
\left[\matrix
{ - & 7 & - & - & 7 & 8 & - & - & - & 6 & - & 3 & 6 & - & - & - & - & - & 
6 & 3 & 8 & 6 & - & - & - & - & 1 & - & - & - \cr  & - & 7 & - & - & 6 & - 
& - & - & 3 & - & 6 & 8 & - & - & 3 & 6 & - & - & - & - & - & - & - & 1 & 
8 & - & - & - & 6 \cr  &  & - & 7 & - & - & - & 6 & 3 & - & - & - & - & 1 & 
- & 6 & 8 & - & - & - & - & - & 6 & 8 & - & 6 & - & - & - & 3 \cr  &  &  & 
- & 7 & - & - & 8 & 6 & - & 6 & - & - & - & 8 & - & - & 1 & - & - & - & - & 
3 & 6 & - & - & - & 6 & 3 & - \cr  &  &  &  & - & - & 1 & - & - & - & 3 & 
- & - & - & 6 & - & - & - & 8 & 6 & 6 & 3 & - & - & - & - & - & 8 & 6 & 
- \cr  &  &  &  &  & - & 7 & - & - & 7 & 8 & - & - & - & 6 & - & 3 & 6 & - 
& - & - & - & - & 6 & 3 & - & - & 1 & - & - \cr  &  &  &  &  &  & - & 7 & 
- & - & 6 & - & - & - & 3 & - & 6 & 8 & - & - & 3 & 6 & - & - & - & 6 & 8 & 
- & - & - \cr  &  &  &  &  &  &  & - & 7 & - & - & - & 6 & 3 & - & - & - & 
- & 1 & - & 6 & 8 & - & - & - & 3 & 6 & - & - & - \cr  &  &  &  &  &  &  &  
& - & 7 & - & - & 8 & 6 & - & 6 & - & - & - & 8 & - & - & 1 & - & - & - & 
- & - & 6 & 3 \cr  &  &  &  &  &  &  &  &  & - & - & 1 & - & - & - & 3 & - 
& - & - & 6 & - & - & - & 8 & 6 & - & - & - & 8 & 6 \cr  &  &  &  &  &  &  & 
 &  &  & - & 7 & - & - & 7 & 8 & - & - & - & 6 & - & 3 & 6 & - & - & - & - 
& - & 1 & - \cr  &  &  &  &  &  &  &  &  &  &  & - & 7 & - & - & 6 & - & - 
& - & 3 & - & 6 & 8 & - & - & - & 6 & 8 & - & - \cr  &  &  &  &  &  &  &  &  
&  &  &  & - & 7 & - & - & - & 6 & 3 & - & - & - & - & 1 & - & - & 3 & 6 & 
- & - \cr  &  &  &  &  &  &  &  &  &  &  &  &  & - & 7 & - & - & 8 & 6 & - 
& 6 & - & - & - & 8 & 3 & - & - & - & 6 \cr  &  &  &  &  &  &  &  &  &  &  & 
 &  &  & - & - & 1 & - & - & - & 3 & - & - & - & 6 & 6 & - & - & - & 8 \cr  
&  &  &  &  &  &  &  &  &  &  &  &  &  &  & - & 7 & - & - & 7 & 8 & - & - & 
- & 6 & - & - & - & - & 1 \cr  &  &  &  &  &  &  &  &  &  &  &  &  &  &  &  
& - & 7 & - & - & 6 & - & - & - & 3 & - & - & 6 & 8 & - \cr  &  &  &  &  &  
&  &  &  &  &  &  &  &  &  &  &  & - & 7 & - & - & - & 6 & 3 & - & - & - & 
3 & 6 & - \cr  &  &  &  &  &  &  &  &  &  &  &  &  &  &  &  &  &  & - & 7 & 
- & - & 8 & 6 & - & 6 & 3 & - & - & - \cr  &  &  &  &  &  &  &  &  &  &  &  
&  &  &  &  &  &  &  & - & - & 1 & - & - & - & 8 & 6 & - & - & - \cr  &  &  
&  &  &  &  &  &  &  &  &  &  &  &  &  &  &  &  &  & - & 7 & - & - & 7 & 1 
& - & - & - & - \cr  &  &  &  &  &  &  &  &  &  &  &  &  &  &  &  &  &  &  & 
 &  & - & 7 & - & - & - & - & - & 6 & 8 \cr  &  &  &  &  &  &  &  &  &  &  & 
 &  &  &  &  &  &  &  &  &  &  & - & 7 & - & - & - & - & 3 & 6 \cr  &  &  &  
&  &  &  &  &  &  &  &  &  &  &  &  &  &  &  &  &  &  &  & - & 7 & - & 6 & 
3 & - & - \cr  &  &  &  &  &  &  &  &  &  &  &  &  &  &  &  &  &  &  &  &  & 
 &  &  & - & - & 8 & 6 & - & - \cr  &  &  &  &  &  &  &  &  &  &  &  &  &  & 
 &  &  &  &  &  &  &  &  &  &  & - & 7 & - & - & 7 \cr  &  &  &  &  &  &  &  
&  &  &  &  &  &  &  &  &  &  &  &  &  &  &  &  &  &  & - & 7 & - & - \cr  & 
 &  &  &  &  &  &  &  &  &  &  &  &  &  &  &  &  &  &  &  &  &  &  &  &  &  
& - & 7 & - \cr  &  &  &  &  &  &  &  &  &  &  &  &  &  &  &  &  &  &  &  &  
&  &  &  &  &  &  &  & - & 7 \cr  &  &  &  &  &  &  &  &  &  &  &  &  &  &  
&  &  &  &  &  &  &  &  &  &  &  &  &  &  & - \cr  } 
\right]
}
$$
\begin{equation} {\cal C}\left(5,\frac 12\right)^{\rm b}=
 \quad\quad\quad\quad\quad\quad \quad\quad\quad\quad\quad\quad\quad
 \quad\quad\quad\quad\quad\quad \quad\quad\quad\quad\quad\quad
\end{equation}
$$
{\tiny
\left[\matrix
{ - & - & 2 & 2 & - & - & - & 9 & 5 & - & - & - & - & 4 & 5 & - & 5 & 4 & 
- & - & - & - & 5 & 9 & - & - & - & - & 4 & 4 \cr - & - & - & 2 & 2 & - & 
4 & 5 & - & - & 5 & - & - & - & 9 & - & - & 4 & 5 & - & - & 4 & 4 & - & - & 
- & - & 9 & 5 & - \cr 2 & - & - & - & 2 & 5 & 4 & - & - & - & 4 & 4 & - & 
- & - & - & - & - & 9 & 5 & 9 & 5 & - & - & - & - & 4 & 5 & - & - \cr 2 & 
2 & - & - & - & 9 & - & - & - & 5 & - & 5 & 9 & - & - & 4 & - & - & - & 4 & 
5 & - & - & - & 4 & 5 & 4 & - & - & - \cr - & 2 & 2 & - & - & - & - & - & 
4 & 4 & - & - & 5 & 4 & - & 5 & 9 & - & - & - & - & - & - & 5 & 4 & 9 & - & 
- & - & 5 \cr - & - & 5 & 9 & - & - & - & 2 & 2 & - & - & - & 9 & 5 & - & 
- & - & - & 4 & 5 & - & 5 & 4 & - & - & 4 & - & - & - & 4 \cr - & 4 & 4 & 
- & - & - & - & - & 2 & 2 & - & 4 & 5 & - & - & 5 & - & - & - & 9 & - & - & 
4 & 5 & - & - & - & - & 9 & 5 \cr 9 & 5 & - & - & - & 2 & - & - & - & 2 & 
5 & 4 & - & - & - & 4 & 4 & - & - & - & - & - & - & 9 & 5 & - & - & 4 & 5 & 
- \cr 5 & - & - & - & 4 & 2 & 2 & - & - & - & 9 & - & - & - & 5 & - & 5 & 
9 & - & - & 4 & - & - & - & 4 & - & 5 & 4 & - & - \cr - & - & - & 5 & 4 & 
- & 2 & 2 & - & - & - & - & - & 4 & 4 & - & - & 5 & 4 & - & 5 & 9 & - & - & 
- & 5 & 9 & - & - & - \cr - & 5 & 4 & - & - & - & - & 5 & 9 & - & - & - & 
2 & 2 & - & - & - & 9 & 5 & - & - & - & - & 4 & 5 & 4 & 4 & - & - & - \cr 
- & - & 4 & 5 & - & - & 4 & 4 & - & - & - & - & - & 2 & 2 & - & 4 & 5 & - & 
- & 5 & - & - & - & 9 & 5 & - & - & - & 9 \cr - & - & - & 9 & 5 & 9 & 5 & 
- & - & - & 2 & - & - & - & 2 & 5 & 4 & - & - & - & 4 & 4 & - & - & - & - & 
- & - & 4 & 5 \cr 4 & - & - & - & 4 & 5 & - & - & - & 4 & 2 & 2 & - & - & 
- & 9 & - & - & - & 5 & - & 5 & 9 & - & - & - & - & 5 & 4 & - \cr 5 & 9 & 
- & - & - & - & - & - & 5 & 4 & - & 2 & 2 & - & - & - & - & - & 4 & 4 & - & 
- & 5 & 4 & - & - & 5 & 9 & - & - \cr - & - & - & 4 & 5 & - & 5 & 4 & - & 
- & - & - & 5 & 9 & - & - & - & 2 & 2 & - & - & - & 9 & 5 & - & - & 4 & 4 & 
- & - \cr 5 & - & - & - & 9 & - & - & 4 & 5 & - & - & 4 & 4 & - & - & - & 
- & - & 2 & 2 & - & 4 & 5 & - & - & 9 & 5 & - & - & - \cr 4 & 4 & - & - & 
- & - & - & - & 9 & 5 & 9 & 5 & - & - & - & 2 & - & - & - & 2 & 5 & 4 & - & 
- & - & 5 & - & - & - & 4 \cr - & 5 & 9 & - & - & 4 & - & - & - & 4 & 5 & 
- & - & - & 4 & 2 & 2 & - & - & - & 9 & - & - & - & 5 & - & - & - & 5 & 
4 \cr - & - & 5 & 4 & - & 5 & 9 & - & - & - & - & - & - & 5 & 4 & - & 2 & 
2 & - & - & - & - & - & 4 & 4 & - & - & 5 & 9 & - \cr - & - & 9 & 5 & - & 
- & - & - & 4 & 5 & - & 5 & 4 & - & - & - & - & 5 & 9 & - & - & - & 2 & 2 & 
- & - & - & 4 & 4 & - \cr - & 4 & 5 & - & - & 5 & - & - & - & 9 & - & - & 
4 & 5 & - & - & 4 & 4 & - & - & - & - & - & 2 & 2 & - & 9 & 5 & - & - \cr 
5 & 4 & - & - & - & 4 & 4 & - & - & - & - & - & - & 9 & 5 & 9 & 5 & - & - & 
- & 2 & - & - & - & 2 & 4 & 5 & - & - & - \cr 9 & - & - & - & 5 & - & 5 & 
9 & - & - & 4 & - & - & - & 4 & 5 & - & - & - & 4 & 2 & 2 & - & - & - & 4 & 
- & - & - & 5 \cr - & - & - & 4 & 4 & - & - & 5 & 4 & - & 5 & 9 & - & - & 
- & - & - & - & 5 & 4 & - & 2 & 2 & - & - & - & - & - & 5 & 9 \cr - & - & 
- & 5 & 9 & 4 & - & - & - & 5 & 4 & 5 & - & - & - & - & 9 & 5 & - & - & - & 
- & 4 & 4 & - & - & - & 2 & 2 & - \cr - & - & 4 & 4 & - & - & - & - & 5 & 
9 & 4 & - & - & - & 5 & 4 & 5 & - & - & - & - & 9 & 5 & - & - & - & - & - & 
2 & 2 \cr - & 9 & 5 & - & - & - & - & 4 & 4 & - & - & - & - & 5 & 9 & 4 & 
- & - & - & 5 & 4 & 5 & - & - & - & 2 & - & - & - & 2 \cr 4 & 5 & - & - & 
- & - & 9 & 5 & - & - & - & - & 4 & 4 & - & - & - & - & 5 & 9 & 4 & - & - & 
- & 5 & 2 & 2 & - & - & - \cr 4 & - & - & - & 5 & 4 & 5 & - & - & - & - & 
9 & 5 & - & - & - & - & 4 & 4 & - & - & - & - & 5 & 9 & - & 2 & 2 & - & 
- \cr  } \right].
}
$$

Finally,
\newpage
\begin{equation} {\cal C}\left(5,\frac 32\right)=
 \quad\quad\quad\quad\quad\quad \quad\quad\quad\quad\quad\quad\quad
 \quad\quad\quad\quad\quad\quad \quad\quad\quad\quad\quad\quad\quad
\end{equation}
$$
{\tiny
\left[\matrix
{ - & 2 & 2 & 2 & 2 & 3 & 3 & 3 & 2 & 3 & 3 & 3 & 3 & 3 & 2 & 2 & 1 & 1 & 
1 & 1 & 2 & 1 & 1 & 1 & 1 & 2 & 2 & 2 & 2 & 4 \cr  & - & 3 & 2 & 3 & 2 & 2 
& 2 & 3 & 1 & 3 & 2 & 1 & 3 & 1 & 3 & 3 & 1 & 1 & 2 & 1 & 2 & 3 & 2 & 2 & 
2 & 1 & 4 & 1 & 2 \cr  &  & - & 3 & 2 & 2 & 2 & 1 & 3 & 2 & 2 & 3 & 3 & 1 & 
3 & 1 & 1 & 3 & 2 & 1 & 1 & 3 & 2 & 2 & 2 & 1 & 2 & 1 & 4 & 2 \cr  &  &  & 
- & 3 & 1 & 3 & 3 & 1 & 2 & 2 & 1 & 2 & 2 & 3 & 1 & 2 & 2 & 2 & 3 & 3 & 1 & 
2 & 1 & 3 & 4 & 1 & 2 & 1 & 2 \cr  &  &  &  & - & 3 & 1 & 2 & 1 & 3 & 1 & 
2 & 2 & 2 & 1 & 3 & 2 & 2 & 3 & 2 & 3 & 2 & 1 & 3 & 1 & 1 & 4 & 1 & 2 & 
2 \cr  &  &  &  &  & - & 3 & 1 & 2 & 2 & 1 & 2 & 1 & 2 & 3 & 1 & 3 & 2 & 3 
& 2 & 2 & 3 & 2 & 1 & 4 & 3 & 1 & 2 & 2 & 1 \cr  &  &  &  &  &  & - & 2 & 
2 & 1 & 2 & 1 & 2 & 1 & 1 & 3 & 2 & 3 & 2 & 3 & 2 & 2 & 3 & 4 & 1 & 1 & 3 & 
2 & 2 & 1 \cr  &  &  &  &  &  &  & - & 3 & 2 & 1 & 3 & 1 & 2 & 2 & 2 & 3 & 
2 & 3 & 1 & 1 & 4 & 2 & 2 & 3 & 1 & 2 & 2 & 3 & 1 \cr  &  &  &  &  &  &  &  
& - & 3 & 1 & 1 & 2 & 2 & 2 & 2 & 2 & 2 & 3 & 3 & 4 & 1 & 1 & 2 & 2 & 3 & 
3 & 1 & 1 & 2 \cr  &  &  &  &  &  &  &  &  & - & 3 & 1 & 2 & 1 & 2 & 2 & 2 
& 3 & 1 & 3 & 1 & 2 & 4 & 3 & 2 & 2 & 1 & 3 & 2 & 1 \cr  &  &  &  &  &  &  & 
 &  &  & - & 2 & 1 & 2 & 2 & 2 & 3 & 2 & 4 & 2 & 3 & 3 & 1 & 2 & 3 & 2 & 3 
& 1 & 2 & 1 \cr  &  &  &  &  &  &  &  &  &  &  & - & 2 & 1 & 2 & 2 & 2 & 3 
& 2 & 4 & 3 & 1 & 3 & 3 & 2 & 3 & 2 & 2 & 1 & 1 \cr  &  &  &  &  &  &  &  &  
&  &  &  & - & 3 & 1 & 3 & 4 & 1 & 3 & 2 & 2 & 3 & 2 & 2 & 3 & 2 & 2 & 3 & 
1 & 1 \cr  &  &  &  &  &  &  &  &  &  &  &  &  & - & 3 & 1 & 1 & 4 & 2 & 3 
& 2 & 2 & 3 & 3 & 2 & 2 & 2 & 1 & 3 & 1 \cr  &  &  &  &  &  &  &  &  &  &  & 
 &  &  & - & 4 & 3 & 1 & 2 & 2 & 2 & 2 & 2 & 3 & 1 & 1 & 3 & 3 & 1 & 2 \cr  
&  &  &  &  &  &  &  &  &  &  &  &  &  &  & - & 1 & 3 & 2 & 2 & 2 & 2 & 2 & 
1 & 3 & 3 & 1 & 1 & 3 & 2 \cr  &  &  &  &  &  &  &  &  &  &  &  &  &  &  &  
& - & 3 & 1 & 2 & 2 & 1 & 2 & 2 & 1 & 2 & 2 & 1 & 3 & 3 \cr  &  &  &  &  &  
&  &  &  &  &  &  &  &  &  &  &  & - & 2 & 1 & 2 & 2 & 1 & 1 & 2 & 2 & 2 & 
3 & 1 & 3 \cr  &  &  &  &  &  &  &  &  &  &  &  &  &  &  &  &  &  & - & 2 & 
1 & 1 & 3 & 2 & 1 & 2 & 1 & 3 & 2 & 3 \cr  &  &  &  &  &  &  &  &  &  &  &  
&  &  &  &  &  &  &  & - & 1 & 3 & 1 & 1 & 2 & 1 & 2 & 2 & 3 & 3 \cr  &  &  
&  &  &  &  &  &  &  &  &  &  &  &  &  &  &  &  &  & - & 3 & 3 & 2 & 2 & 1 
& 1 & 3 & 3 & 2 \cr  &  &  &  &  &  &  &  &  &  &  &  &  &  &  &  &  &  &  & 
 &  & - & 2 & 2 & 1 & 3 & 2 & 2 & 1 & 3 \cr  &  &  &  &  &  &  &  &  &  &  & 
 &  &  &  &  &  &  &  &  &  &  & - & 1 & 2 & 2 & 3 & 1 & 2 & 3 \cr  &  &  &  
&  &  &  &  &  &  &  &  &  &  &  &  &  &  &  &  &  &  &  & - & 3 & 3 & 1 & 
2 & 2 & 3 \cr  &  &  &  &  &  &  &  &  &  &  &  &  &  &  &  &  &  &  &  &  & 
 &  &  & - & 1 & 3 & 2 & 2 & 3 \cr  &  &  &  &  &  &  &  &  &  &  &  &  &  & 
 &  &  &  &  &  &  &  &  &  &  & - & 3 & 2 & 3 & 2 \cr  &  &  &  &  &  &  &  
&  &  &  &  &  &  &  &  &  &  &  &  &  &  &  &  &  &  & - & 3 & 2 & 2 \cr  & 
 &  &  &  &  &  &  &  &  &  &  &  &  &  &  &  &  &  &  &  &  &  &  &  &  &  
& - & 3 & 2 \cr  &  &  &  &  &  &  &  &  &  &  &  &  &  &  &  &  &  &  &  &  
&  &  &  &  &  &  &  & - & 2 \cr  &  &  &  &  &  &  &  &  &  &  &  &  &  &  
&  &  &  &  &  &  &  &  &  &  &  &  &  &  & - \cr  } 
\right].
}
$$

\section*{Acknowledgments}

We are indebted to M.\ Wierzbowska, G.\ Santoro, E.\ Tosatti
for useful discussions.
This work was supported by the European Union, contract
ERBFMRXCT970155 (TMR FULPROP), covering in particular the 
postdoctoral work of M. L{\"u}ders, and by MURST COFIN01.



\end{document}